\documentclass[conference]{IEEEtran}
\IEEEoverridecommandlockouts
\usepackage{algorithm}
\usepackage{algpseudocode}
\usepackage{pgfplots}
\pgfplotsset{compat=1.18}

\usepackage{cite}
\usepackage{amsmath,amssymb,amsfonts}
\usepackage{graphicx}
\usepackage{textcomp}
\usepackage{xcolor}
\usepackage[T1]{fontenc}
\usepackage{lmodern}

\usepackage{float}

\usepackage{tikz}
\usetikzlibrary{positioning}

\usepackage{longtable}
\usepackage{booktabs}
\usepackage{array}
\usepackage{tabularx}
\usepackage{enumitem}

\usepackage{hyperref}
\usepackage[nameinlink]{cleveref}

\usepackage{listings}
\usepackage{verbatim}

\usepackage{multirow}
\usepackage{makecell}
\newcommand{\entity}[3]{\textbf{#2}}
\def\BibTeX{{\rm B\kern-.05em{\sc i\kern-.025em b}\kern-.08em
    T\kern-.1667em\lower.7ex\hbox{E}\kern-.125emX}}
    
\usepackage{microtype}
\begin{document}

\title{Towards a Bridge Layer Between Bibliographic and Formalized Mathematical Knowledge}

\author{\IEEEauthorblockN{ }
\IEEEauthorblockA{Arnaud Mayeux \\
\textit{University of Wisconsin-Madison}\\
 \\
}}

\maketitle

\begin{abstract}
Mathematical knowledge is split between bibliographic databases (e.g., MathSciNet, zbMATH Open) and formal proof libraries (e.g., Lean’s mathlib), preventing unified access between published results and their formalizations. We propose a relational bridge-database that aligns publication metadata with formal artifacts, providing an interoperability layer between mathematical literature and machine-verifiable proofs. We introduce a paper-level formalization score that measures how much of a publication is covered in formal systems. As a feasibility study, we show how such scores can be estimated via cross-document alignment between informal texts and Lean formalizations, enabling large-scale analysis of formalization coverage. This framework is a first step toward integrating bibliographic and formal mathematical ecosystems into scalable, machine-actionable knowledge graphs linking publications to formal proof objects.
\end{abstract}


\section{Introduction}
The architecture of modern mathematical knowledge is characterized by a fundamental dichotomy. On one hand, the vast corpus of human-readable mathematical research is indexed by bibliographic databases such as MathSciNet~\cite{mathscinet} and zbMATH Open~\cite{zbmath}, which capture the publication record, citation structure, and author networks. On the other hand, interactive theorem provers (ITPs) such as Lean and libraries like \textit{mathlib}~\cite{mathlib} provide machine-verifiable mathematics encoded as formally checked objects.

Despite their complementary roles, these infrastructures remain largely disconnected. Bibliographic systems do not encode whether results have been formalized, while formal libraries are organized around internal dependency graphs rather than publication metadata.

In practice, however, the boundary between these ecosystems is partially bridged by existing “alignment artifacts.” Formalization projects, library documentation, and survey papers often explicitly describe how classical results correspond to their formal counterparts, effectively forming informal dictionaries between publications and mechanized developments. These mappings exist but remain fragmented and non-standardized across sources.

This observation motivates not a replacement of existing infrastructures, but a systematization of these partial correspondences into a unified relational layer. The goal is to turn existing informal alignment knowledge into a structured, queryable bridge between bibliographic databases and formal mathematical libraries.

In this paper, we introduce a relational bridge-database integrating bibliographic registries with formal verification environments. The contributions are three-fold:
\begin{itemize}
\item We propose an interoperability layer linking MathSciNet and zbMATH Open records with formal artifacts in systems such as Lean.
\item We define a paper-level formalization score $S(P_i)$ measuring the extent of alignment between published mathematical statements and formal counterparts.
\item We present a feasibility study showing how alignment signals from existing formalization artifacts can be used to estimate coverage in a consistent scoring framework.
\end{itemize}

The remainder of the paper is organized as follows. Section II describes current limitations and the proposed architecture. Section III introduces the scoring procedure. Section IV presents case studies. Sections V and VI discuss scalability and limitations.

\section{Current databases and the proposed new bridge-database}

Modern mathematical knowledge is organized across two largely independent infrastructures: bibliographic databases that index published research mathematics, and formal systems that encode machine-verifiable mathematics. In addition, there already exists a growing ecosystem of formalized mathematical databases; however, these are not systematically connected to the research-level publication graph.
In practice, the relationship between informal mathematical publications and their formal counterparts is often already partially documented by the mathematical community itself. Survey papers, formalization projects, and library documentation frequently act as informal dictionaries that describe how classical results correspond to mechanized developments in proof assistants. These mappings are not incidental; they are a natural byproduct of the formalization workflow, where authors explicitly record how definitions, theorems, and constructions from the literature are represented in systems such as Lean. Our goal is therefore not to infer entirely unknown correspondences in a vacuum, but to systematize and scale this existing practice into a machine-actionable relational layer that can be queried, quantified, and updated across the mathematical corpus.

\subsection{Bibliographic databases of mathematical research}
The primary infrastructure for indexing published mathematics is \textbf{\entity{software}{MathSciNet}{bibliographic database of mathematical literature}} \cite{mathscinet}, which catalogs peer-reviewed journal articles that form the canonical research record. It operates within the traditional publication paradigm, where results are validated through editorial processes and peer review across journals of varying prestige. As a comprehensive historical repository, MathSciNet indexes over 4.5 million entries spanning pure and applied mathematics.

Each entry in MathSciNet is embedded in a rich citation network, linking references, citations, journal metadata, author information, and often expert-written mathematical reviews. This makes it both a bibliographic index and a curated evaluative layer over the mathematical literature.

A closely related system is \textbf{\entity{software}{zbMATH Open}{mathematics abstracting and reviewing database}} \cite{zbmath}, which provides similar indexing functionality and additionally includes broader coverage of preprints and auxiliary metadata.

\subsection{Existing formalized mathematics databases}

In parallel to the publication ecosystem, there already exist formalized mathematics databases built within proof assistant environments. These systems store mathematics as fully machine-checked objects rather than human-readable publications.

The most developed example is \textbf{\entity{software}{mathlib}{Lean mathematical library}} \cite{mathlib}, which forms a large-scale formal encyclopedia of mathematics inside the Lean proof assistant. Its contents are not derived from journal authority but from internal logical correctness verified by a kernel.

More broadly, repositories such as \textbf{\entity{software}{Reservoir}{formalization project repository connected to Lean ecosystems}} \cite{reservoir} collect formalization projects that extend beyond curated libraries. These systems already constitute a form of formal mathematics database, but they are primarily organized around proof-assistant development rather than alignment with the published research literature.

Formally verified mathematics and large formal libraries have been developed over many years \cite{Boy94, kohlhase} and are regularly discussed in the literature \cite{Blan, avigad2014, Mayeux2026Indexed}. Formally verified mathematics provides an essential framework for fully certified AI-generated mathematics, as well as for large-scale human–AI hybrid mathematical work.

\subsection{Structural disconnect between databases}

There is currently no unified infrastructure that connects research mathematics databases (such as MathSciNet or zbMATH Open) with formal systems (such as mathlib or Reservoir).

This disconnect persists despite the fact that formalized mathematics databases already exist and are actively growing. However, these databases are designed within the paradigm of formalized mathematics: they organize mathematical knowledge from the perspective of what has been formalized. These systems are developed independently of the classical publication ecosystem and are not aligned with journal metadata, citation graphs, or publication hierarchies.

As a result, formalized mathematics and published mathematics constitute two partially overlapping but structurally independent representations of mathematical knowledge.

\subsection{Motivation: need for a bridge rather than replacement}

The goal is not to replace existing databases. Bibliographic systems and formal libraries serve fundamentally different roles:

\begin{itemize}
\item bibliographic databases encode the human research record
\item formal systems encode machine-verifiable mathematical content
\end{itemize}

What is missing is a connective layer between them.

\subsection{Proposed bridge-database}

We therefore propose a bridge-database that integrates existing infrastructures without replacing them. The database includes all papers indexed in \textbf{\entity{software}{MathSciNet}{bibliographic database of mathematical literature}} and associates to each paper:

\begin{itemize}
\item a formalization score, measuring the extent of formalization
\item links to:
  \begin{itemize}
  \item MathSciNet metadata
  \item zbMATH Open
  \item the publishing journal
  \item formalization code (if available)
  \item an optional mapping between PDF content and formalization
  \end{itemize}
\end{itemize}

The formalization score is determined in an elementary manner from data provided by authoritative institutions, such as journals dedicated to formalized and informal mathematics. Its definition is introduced below.

\begin{figure}[!ht]
\centering
\begin{tikzpicture}[
    node distance=0.45cm,
    box/.style={
        draw,
        rounded corners,
        text width=3.0cm,
        align=center,
        font=\footnotesize,
        inner sep=4pt
    },
    >=latex
]

\node[box] (bib) {
\textbf{Bibliographic databases}\\
MathSciNet\\
zbMATH Open\\
arXiv\\
Journal metadata
};

\node[box, below=of bib] (bridge) {
\textbf{Bridge database}\\
Formalization scores\\
Statement mappings\\
Metadata links\\
Cross-references
};

\node[box, below=of bridge] (formal) {
\textbf{Formal mathematics databases}\\
mathlib\\
Reservoir\\
Lean projects\\
Formal theorem objects
};
\draw[->, thick] (bridge) -- (bib);
\draw[->, thick] (bridge) -- (formal);

\end{tikzpicture}
\caption{Bridge architecture connecting bibliographic databases with formal mathematics databases through an interoperability layer.}
\label{fig:bridge}
\end{figure}
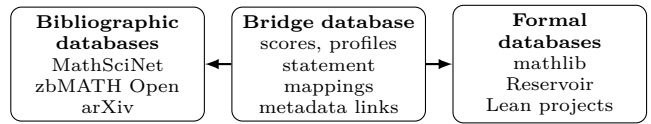

The proposed system is not a replacement for existing publication infrastructures; rather, it serves as an alignment infrastructure. The new database has no validation function, either at the peer-review stage or at the level of formal compilation, and it does not replace journals dedicated to informal or formalized mathematics. Instead, it acts as an archival platform that preserves and interconnects their contributions. Figure~\ref{fig:bridge} shows the proposed bridge architecture.

\section{Paper-Level Formalization Scoring}

We treat each indexed paper $P_i$ as a structured collection of mathematical statements (definitions, lemmas, theorems, propositions, constructions, and conjectures). Each statement is considered uniformly, without manual weighting.

For a paper $P_i$, let:
\begin{itemize}
    \item $\mathcal{S}_i$ be the set of all mathematical statements in the paper,
    \item $\mathcal{F}_i \subset \mathcal{S}_i$ be the subset of statements that are formalized in Lean.
\end{itemize}

A statement $s$ is considered formalized only if all required proof obligations are also formalized whenever $s$ is of a provable type (theorem, lemma, proposition, etc.).

We define the binary formalization indicator:
\[
\phi(s) =
\begin{cases}
1 & \text{if $s$ is fully formalized in Lean}, \\
0 & \text{otherwise}.
\end{cases}
\]

\subsection{Statement Categories}

We distinguish two regimes depending on the nature of each statement:

\begin{itemize}
    \item A \textbf{pure statement} (e.g. definition, construction, conjecture) requires only the existence of its formal counterpart in Lean (no proof needed).
    \item A \textbf{proved statement} (e.g. theorem, lemma, proposition) requires both the formal statement \emph{and} a complete formal proof in Lean.
\end{itemize}

\subsection{Formalization Scoring Procedure}

The formalization score of a paper is computed as follows:

\begin{enumerate}
    \item Extract all mathematical statements from the paper.
    \item For each statement $s \in \mathcal{S}_i$, search for a corresponding formalization in Lean.
\end{enumerate}

A statement contributes to the score if it satisfies its category-specific condition:
\begin{itemize}
    \item Pure statements (definitions, constructions, conjectures): the formal statement exists in Lean.
    \item Proved statements (theorems, lemmas, propositions): both the statement and its proof exist in Lean.
\end{itemize}

The score is then:
\[
S(P_i) = 100 \cdot \frac{\sum_{s \in \mathcal{S}_i} \phi(s)}{|\mathcal{S}_i|},
\]
with the convention that $S(P_i) = 0$ if $\mathcal{S}_i = \emptyset$.

In general, the formalization score may be computed using multiple formalization sources corresponding to a given paper; however, for simplicity, we restrict this proof-of-concept study to a single reference formalization document (PDF2) for each evaluated paper.

\section{Case Studies}

We conducted experiments on several mathematical papers in PDF format to evaluate whether a large language model can independently compute paper-level formalization scores. The experiments were performed using Google Gemini (University of Wisconsin–Madison access), without any task-specific fine-tuning.

For each evaluation, the model received a fixed three-part input consisting of: (i) the instruction prompt defined in Section~\ref{ssi}, (ii) PDF1, the mathematical paper under analysis, and (iii) PDF2, the corresponding formalization document in Lean (or a structured description of its Lean implementation).

All runs were executed in a zero-shot setting to ensure reproducibility. The model produced a single-pass output for each paper pair, without iterative refinement or human intervention.

We additionally performed manual verification of the generated scores and alignment results by inspecting the corresponding mathematical papers and formalization documents ourselves. The manually assessed results closely matched the LLM-generated estimates, providing an initial qualitative validation of the approach.

We additionally provide a detailed example in Section~\ref{fine} illustrating the alignment process and the computation of the formalization score for a representative mathematical paper.

\subsection{Instruction} \label{ssi}
 \begin{quote}\scriptsize

\ttfamily

\textbf{CROSS-DOCUMENT FORMALIZATION ALIGNMENT TASK}

You are given two documents: PDF 1 (an informal mathematical paper) and PDF 2 (a Lean formalization or related documentation).

Your task is to estimate a paper-level formalization score measuring how much of the mathematical content in PDF 1 is represented in PDF 2.

\textbf{STEP 1 — Identify mathematical content}

Read PDF 1 and identify all primary mathematical statements, including definitions, theorems, lemmas, propositions, constructions, claims, and conjectures.

Strict Segmentation Rule: If a high-level mathematical structure contains explicitly partitioned sub-statements (e.g., "Lemma 1" containing distinct parts "a, b, c, d", or a Theorem listing multiple independent clauses), you must count each individual sub-part as a separate, distinct mathematical statement.

Exclude all surrounding narrative text, scratchpad equations, proof steps, and intermediate derivations. Each standalone claim or sub-claim must be isolated and counted exactly once.

Let $N$ be the total number of these granular, extracted mathematical statements.

\textbf{STEP 2 — Classify statements}

Assign each identified statement to one of two categories:

PURE statements: definitions, constructions, conjectures, or introduced mathematical objects.

PROVED statements: theorems, lemmas, propositions, sub-parts of lemmas/theorems, or claims with proofs.

Classification should be based on the role of the statement in the text, not formatting.

\textbf{STEP 3 — Match to formalization}

For each statement, determine whether PDF 2 contains evidence of a corresponding formalization in Lean.

Evidence may include:

- a Lean definition, theorem, or object with matching mathematical meaning

- a file, module, or comment describing the formalization of the result

- an explicit or implicit mapping between informal and formal content

Use semantic similarity and mathematical equivalence, not exact wording.

If no reasonable correspondence can be identified, mark as not formalized.

\textbf{STEP 4 — Scoring}

PURE statement: score 1 if a Lean formal object exists, otherwise 0.

PROVED statement: score 1 if the statement is formalized in Lean and the proof is also represented or constructively encoded, otherwise 0.

Each statement contributes equally to the score.

Let $M$ be the number of statements scored 1.

\[
S(P) = 100 \cdot \frac{M}{N}
\]

If $N = 0$, output 0.

\textbf{OUTPUT FORMAT}

Return exactly:

(1) a LaTeX table with rows:

Total Statements, Formalized Statements, Formalization Score

(2) a plain-text version of the same table.

Do not include explanations or extra commentary.

\end{quote} 

\subsection{Experiment on several papers}

 \subsubsection{Beukers \cite{beukers1979} : irrationality of $\zeta (2) $ and $\zeta (3)$}
 
The irrationality of $\zeta(3)$ has been formalized in \cite{liu2025}. In this experiment, we consider PDF1~\cite{beukers1979} and PDF2~\cite{liu2025}.
 
 \begin{table}[ht]
\centering
\caption{Model-generated formalization scoring results for Irrationality of $\zeta (3)$}
\begin{tabular}{|l|l|}
\hline
Quantity & Value \\ \hline
Total Statements Counted & 6 \\ \hline
Total Statements Formalized & 4 \\ \hline
Formalization Score & 66.67 \\ \hline
\end{tabular}
\end{table}

The model output indicates that the formalization of \cite{beukers1979} has been substantially achieved, though not all propositions concerning the mathematical content in PDF 1 have been implemented in PDF 2. Interested readers consulting the proposed database may examine which parts have been formalized and which have not.

\subsubsection{The sphere packing problem in dimension 8}

The sphere packing problem in dimension 8 was solved in 2016 \cite{Via17}, representing a major breakthrough in discrete geometry. A formalization of this result in Lean was completed in 2026 \cite{SpherePacking8}.

In this experiment, PDF1 corresponds to the original mathematical exposition of the 8-dimensional sphere packing solution \cite{Via17}, while PDF2 corresponds to its Lean formalization documentation as described in \cite{SpherePacking8}.

\begin{table}[ht]
\centering
\caption{Model-generated formalization scoring results for sphere packing in dimension 8}
\begin{tabular}{|l|l|}
\hline
Quantity & Value \\ \hline
Total Statements Counted & 12 \\ \hline
Total Statements Formalized & 12 \\ \hline
Formalization Score & 100 \\ \hline
\end{tabular}
\end{table}

The model output indicates that formalization has been achieved.

\subsubsection{Dilatations of categories}

Dilatations of categories is a general, recently published construction in categorical algebra \cite{MayDil}. A partial formalization of this theory was described in \cite{Mayeux2026Indexed}.

In this experiment, PDF1 corresponds to \cite{MayDil}, and PDF2 corresponds to \cite{Mayeux2026Indexed}.
\begin{table}[ht]
\centering
\caption{Model-generated formalization scoring results for dilatations of categories}
\begin{tabular}{|l|l|}
\hline
Quantity & Value \\ \hline
Total Statements Counted & 48 \\ \hline
Total Statements Formalized & 20 \\ \hline
Formalization Score & 41.67 \\ \hline
\end{tabular}
\end{table}

The model output indicates that the formalization has been only partly done.

\subsubsection{Multi-graded Proj schemes}

Multi-graded Proj schemes is a general construction in algebraic geometry \cite{MR26}. The paper \cite{MR26} can be divided into two parts: the first part contains the foundations of the theory, while the second part concerns applications.

Part 1 was (partially) formalized in \cite{MZ26} (cf. also \cite{MZproc}).

In this experiment, PDF1 corresponds to \cite[Pages 1–16]{MR26} and PDF2 corresponds to \cite{MZ26}.
\begin{table}[ht]
\centering
\caption{Model-generated formalization scoring results for multigraded Proj schemes}
\begin{tabular}{|l|l|}
\hline
Quantity & Value \\ \hline
Total Statements Counted & 31 \\ \hline
Total Statements Formalized & 21 \\ \hline
Formalization Score & 67.74 \\ \hline
\end{tabular}
\end{table}

The model output indicates that the formalization has been substantially achieved, though not all propositions concerning properties of Proj have yet been implemented.

\subsubsection{Multi-graded Proj schemes, second experiment}

We tested the experiment with PDF1=PDF2 being \cite{MZ26}.
\begin{table}[ht]
\centering
\caption{Model-generated formalization scoring results for multigraded Proj schemes, 2}
\begin{tabular}{|l|l|}\hline
Quantity & Value \\ \hline
Total Statements Counted & 54 \\ \hline
Total Statements Formalized & 52 \\ \hline
Formalization Score & 96.30 \\ \hline
\end{tabular}
\end{table}

The model output indicates a very high formalization rate; the difference may be due to the informal phrasing in the presentation paper and the use of external references.

\subsubsection{Magnets and attractors of diagonalizable group schemes}

Algebraic magnetism is a framework in dynamical algebraic geometry. It has not yet been implemented in Lean 4.

In this experiment, PDF1 was a survey of this theory \cite{MayAttract}, and PDF2 was an unrelated document \cite{MZ26} (the documentation of the formalization of Proj).

\begin{table}[ht]
\centering
\caption{Model-generated formalization scoring results for algebraic magnetism}
\begin{tabular}{|l|l|}
\hline
Total Statements Counted & 18 \\ \hline
Total Statements Formalized & 0 \\ \hline
Formalization Score & 0 \\ \hline
\end{tabular}
\end{table}

The system detects that PDF2 contains nothing relevant. 

\section{Feasibility of Large-Scale Integration}

The proposed bridge-database is technically feasible because it does not introduce a new evaluation or verification layer. Both mathematical publication and formal verification infrastructures already exist and already perform their respective validation functions.

Mathematical papers are reviewed and curated through established publication systems and indexed by databases such as MathSciNet and zbMATH Open. Similarly, formal mathematical artifacts are validated by proof assistant kernels and maintained within formal libraries such as mathlib and related repositories. The proposed bridge-database does not attempt to replace either process.

Instead, the system functions purely as a relational database that records correspondences between existing entities. For a given paper, the database stores identifiers, metadata, links to bibliographic records, links to formalization projects, and a formalization score derived from available formalization data. No additional mathematical verification is performed by the database itself.

Furthermore, the database can be updated incrementally as formalization coverage evolves over time, without affecting existing publication or verification infrastructures.

The computation of formalization scores also appears feasible in practice. The experiments reported show that a contemporary large language model can successfully compare mathematical papers with corresponding formalization documents and recover meaningful paper-level formalization scores across multiple scenarios, including fully formalized, partially formalized, and unrelated document pairs. Although these experiments are preliminary, they suggest that parts of statement extraction and matching can be automated with LLM assistance. At the end, hyperlinks between papers will be added manually or supervised manually, as is the case with MathSciNet and zbMATH.

\section{Fine-Grained Formalization Mapping for \cite{beukers1979}} \label{fine}

In this section, we provide a detailed verification of the formalization score in the case of the irrationality of $\zeta (2)$ and $\zeta(3)$ \cite{beukers1979}, by comparing the mathematical statements in \cite{beukers1979} with their corresponding Lean 4 implementation in \cite{liu2025}. While the score introduced above provides a global measure of formalization coverage, a more detailed record of the individual correspondences provides additional information about which mathematical components have been formalized and which remain uncovered.

Such statement-level information can naturally be stored in the proposed bridge-database as a refinement of the paper-level score. The database may therefore contain multiple levels of granularity, ranging from a global formalization score associated with a paper, to intermediate summaries, and finally to detailed mappings between individual mathematical statements and their formal counterparts.

\begin{table}[h]
\centering
\caption{Mapping Beukers \cite{beukers1979} to formalization \cite{liu2025}}
\label{tab:formalization_map}
\begin{tabular}{|l|p{3.5cm}|p{2.5cm}|}
\hline
\textbf{Ref in \cite{beukers1979}} & \textbf{Statement in \cite{beukers1979}} & \textbf{Formalization status in \cite{liu2025}} \\
\hline
Lemma 1a & Rationality of $\int \int \frac{x^r y^s}{1-xy}$ & Formalized: Sec 2, 5  \\
\hline
Lemma 1b & Rationality of log-integral & Formalized: Sec 2, 5  \\
\hline
Lemma 1c & Identity for $\zeta(2) - \sum k^{-2}$ & Not formalized  \\
\hline
Lemma 1d & Identity for $2(\zeta(3) - \sum k^{-3})$ & Formalized: Sec 5  \\
\hline
Theorem 1 & Irrationality of $\zeta(2)$ & Not formalized  \\
\hline
Theorem 2 & Irrationality of $\zeta(3)$ & Formalized: Sec 5  \\
\hline
\end{tabular}
\end{table}

\section{Limitations, comments and further directions}

Despite its utility as an alignment layer between bibliographic and formal mathematical infrastructures, the proposed framework has limitations. PDF-based mathematical texts do not encode explicit structural information, which can introduce ambiguity in parsing and reconstructing formal statements. The extraction, classification, and weighting of mathematical statements remain imperfect, particularly when it comes to distinguishing between definitions, remarks, and auxiliary narrative content.

The accuracy of the proposed scoring procedure depends on retrieval quality when matching informal statements to formal artifacts, making it sensitive to indexing and semantic search performance. Large language models may also produce incorrect or spurious correspondences between informal and formal statements in ambiguous cases.

Finally, large-scale indexing systems, including both bibliographic and formal mathematical databases, may contain incomplete, duplicated, or inconsistently curated records, which can introduce noise in cross-database alignment. Overall, these considerations suggest that the proposed system should be regarded as a best-effort framework for relational alignment rather than a fully faithful mapping between the two representations.

Papers are assigned label D for scores below $10\%$, label C for scores between $10\%$ and $50\%$, label B for scores between $50\%$ and $90\%$, and label A for scores between $90\%$ and $100\%$. These labels are intended as an indexing aid for large-scale databases, while the numerical score and detailed statement-level mappings remain available when finer analysis is required.
\begin{table}[ht]
\centering
\caption{Formalization levels of studied papers}
\resizebox{\columnwidth}{!}{
\begin{tabular}{|l|c|c|}
\hline
\textbf{Paper} & \textbf{Score} & \textbf{Level} \\
\hline
Beukers  & 66.67 & B \\
\hline
Sphere packing  & 100 & A \\
\hline
Dilatations of categories & 41.67 & C \\
\hline
Multigraded Proj schemes & 67.74 & B \\
\hline
Multigraded Proj (self) & 96.30 & A \\
\hline
Algebraic magnetism & 0 & D \\
\hline
\end{tabular}
}
\end{table}

As an additional robustness check, we repeated the evaluation using multiple large language models. The resulting formalization scores showed only minor variations, and in our cases all tested models assigned the evaluated papers to the same qualitative categories (A–D).

The formalization score $S(P_i)$ introduced here serves as an initial baseline. To improve sensitivity, we define a weighted formalization score that acknowledges the internal hierarchy of mathematical research:

$$
S(P_i) = 100 \cdot \frac{\sum_{s \in \mathcal{S}_i} \omega(s) \cdot \phi(s)}{\sum_{s \in \mathcal{S}_i} \omega(s)}
$$

where $\phi(s) \in \{0, 1\}$ denotes formalization status, and $\omega(s)$ is a weight assigned based on the statement type: $\omega = 3.0$ for theorems and propositions, $\omega = 2.0$ for lemmas and corollaries, and $\omega = 1.0$ for definitions and auxiliary constructions. This refinement may ensure a more accurate score.

Applying our weighted scoring scheme to \cite{beukers1979} yields a score of $64.3\%$, which maintains the same Label B designation as our classical unweighted score of $66.7\%$. In this proof-of-concept paper, we used the unweighted score as a simpler, objective, and consistent measure of formalization coverage, as weighting provided minimal improvement. It is also not clear that definitions should systematically receive lower weights than theorems or propositions, since foundational definitions may play a central role in the formalization of a theory. Therefore, the choice of statement weights remains application-dependent and requires further investigation.

More generally, the bridge-database supports multiple representation levels, from global paper indicators to detailed mappings between mathematical statements and formal counterparts. Future work will explore richer alignment models, alternative scoring functions, and additional metadata.


\small
\begin{thebibliography}{00}

\bibitem{avigad2014}
J. Avigad and J. Harrison.
\textit{Formally Verified Mathematics}.
Communications of the ACM, 57(4):66--75, 2014.
DOI: 10.1145/2609092

\bibitem{beukers1979}
F. Beukers.
A note on the irrationality of $\zeta(2)$ and $\zeta(3)$.
\textit{Bulletin of the London Mathematical Society}, 11 (1979), 268--272.


\bibitem{Blan} J.~C. Blanchette, C.~Kaliszyk, L.~C. Paulson, and J.~Urban, Hammering towards QED, J. Formaliz. Reason. {\bf 9} (2016), no.~1, 101--148; MR3460643

\bibitem{kohlhase}
M.~Kohlhase and F.~Rabe,
\textit{QED reloaded: towards a pluralistic formal library of mathematical knowledge},
Journal of Formalized Reasoning, \textbf{9} (2016), no.~1, 201--234.

\bibitem{Boy94} Boyer, R.S. (1994). A mechanically proof-checked encyclopedia of mathematics: Should we build one? Can we?. In: Bundy, A. (eds) Automated Deduction — CADE-12. CADE 1994. Lecture Notes in Computer Science, vol 814. Springer, Berlin

\bibitem{kojima2022}
T. Kojima, S. S. Gu, M. Reid, Y. Matsuo, and Y. Iwasawa.
\textit{Large Language Models are Zero-Shot Reasoners}.
In Advances in Neural Information Processing Systems (NeurIPS), vol. 35, pp. 22199--22213, Curran Associates, Inc., 2022.

\bibitem{Lange}
C. Lange.
\newblock Ontologies and Languages for Representing Mathematical Knowledge on the Semantic Web.
\newblock \emph{Semantic Web}, 4(2):119--158, 2013.

\bibitem{minerva2022}
A. Lewkowycz, A. Andreassen, D. Dohan, E. Dyer, H. Michalewski, V. Ramasesh, A. Slone, C. Anil, I. Schlag, T. Gutman-Solo, Y. Wu, B. Neyshabur, G. Gur-Ari, and V. Misra.
\textit{Solving Quantitative Reasoning Problems with Language Models}.
In Advances in Neural Information Processing Systems (NeurIPS), vol. 35, pp. 3843--3857, Curran Associates, Inc., 2022.


\bibitem{liu2025}
J. Liu, J. Zhang, L. Zhi,
A Formal Proof of the Irrationality of $\zeta(3)$ in Lean~4.
\textit{Journal of Systems Science and Complexity}, 2025, XX: 1--30.

\bibitem{MayDil} A. Mayeux, Dilatations of categories, High. Struct. {\bf 9} (2025), no.~2, 62--75.

\bibitem{MayAttract} A. Mayeux, Magnets and attractors of diagonalizable group schemes actions, in {\it Lie theory and its applications in physics}, 437--442, Springer Proc. Math. Stat., 473, Springer, Singapore

\bibitem{Mayeux2026Indexed}
A. Mayeux,
{Formalizing all indexed mathematics as a benchmark for general reasoning, with the example of implementing dilatations of categories},
arXiv:2606.03835, 2026.

\bibitem{MR26} A. Mayeux and S. Riche, On multi-graded Proj schemes, Publ. Res. Inst. Math. Sci. {\bf 62} (2026), no.~1, 115--176

\bibitem{MZproc}
A. Mayeux and J. Zhang,
\textit{The Mechanization of Science Illustrated by the Lean Formalization of the Multi-graded Proj Construction},
Springer Proceedings in Mathematics \& Statistics, to appear.


\bibitem{MZ26}
A. Mayeux and J. Zhang,
\emph{Formalizing multi-graded Brenner-Schröer Proj schemes and dilatations of rings in Lean4},
arXiv:2606.01438, 2026.



\bibitem{mathscinet}
American Mathematical Society (AMS).
\textit{MathSciNet: Mathematical Reviews Database}.

\bibitem{mathlib}
The mathlib Community.
\textit{mathlib: The Lean Mathematical Library}.

\bibitem{reservoir}
The Lean Community.
\textit{Reservoir: The Lean Package Registry and Indexing Infrastructure}.
Available: https://reservoir.lean-lang.org/

\bibitem{hermes2026}
A. Ospanov, Z. Feng, J. Sun, H. Bai, S. Xin, and F. Farnia.
\textit{HERMES: Towards Efficient and Verifiable Mathematical Reasoning in LLMs}.
In Proceedings of the 43rd International Conference on Machine Learning (ICML), PMLR 306, Seoul, South Korea, 2026.

\bibitem{SpherePacking8}
S. Hariharan, C. Birkbeck, S. Lee, H. K. G. Ma, B. Mehta, A. Poiroux, and M. Viazovska,
\emph{Progress in Formalizing Sphere Packing in Dimension 8},
arXiv:2604.23468 [math.MG], 2026.

\bibitem{Via17} M.~S. Viazovska, The sphere packing problem in dimension 8, Ann. of Math. (2) {\bf 185} (2017), no.~3, 991--1015

\bibitem{zbmath}
FIZ Karlsruhe, European Mathematical Society, and Heidelberg Academy of Sciences and Humanities.
\textit{zbMATH Open}.
Mathematical literature database.

\end{thebibliography}
\end{document}